\documentclass[aps,prd,nofootinbib,twocolumn,superscriptaddress,preprintnumbers,letterpaper,natbib,nofootinbib]{revtex4-2}
\pdfoutput = 1
\usepackage{amssymb}
\usepackage{amsmath}
\usepackage{epsfig}
\usepackage{hyperref}
\usepackage{color}
\usepackage{cleveref}
\usepackage{multirow}
\usepackage{capt-of}
\usepackage{siunitx}
\usepackage{graphicx}
\usepackage[caption=false]{subfig}
\usepackage{slashed}
\usepackage{tabularx}
\usepackage{enumitem}
\usepackage{multirow}

\newcommand{\be}{\begin{equation}}
\newcommand{\ee}{\end{equation}}
\newcommand{\bea}{\begin{eqnarray}}
\newcommand{\eea}{\end{eqnarray}}

\makeatletter
\def\simgt{\mathrel{\lower2.5pt\vbox{\lineskip=0pt\baselineskip=0pt
           \hbox{$>$}\hbox{$\sim$}}}}
\def\simlt{\mathrel{\lower2.5pt\vbox{\lineskip=0pt\baselineskip=0pt
           \hbox{$<$}\hbox{$\sim$}}}}
\makeatother

\newcommand{\avg}[1]{\left\langle#1\right\rangle}
\newcommand{\prn}[1]{ \left(  #1 \right) }

\newcommand{\al}[1]{\begin{align} #1 \end{align}}

\newcommand{\dl}{\ell_d}
\newcommand{\dlbar}{\bar{\ell}_d}
\newcommand{\Ydl}{Y_{\dl}}
\newcommand{\YLdark}{Y_L^{\text{dark}}}
\newcommand{\YLSM}{Y_L^{\text{SM}}}
\newcommand{\YBSM}{Y_B^{\text{SM}}}
\newcommand{\YBmeas}{Y_B^{\text{obs}}}
\newcommand{\di}{\chi_1}
\newcommand{\dibar}{\bar{\chi}_1}
\newcommand{\df}{\chi_2}

\newcommand{\BSM}{\mathcal{B}}
\newcommand{\BSMb}{\overline{\mathcal{B}}}
\newcommand{\nBSMnBSMb}{\prn{n_{\BSM}-n_{\BSMb}}}
\newcommand{\sigv}{\avg{\sigma v}}
\newcommand{\ACP}{A_{CP}^f}
\newcommand{\ACPfrac}{a_{CP}^f}
\newcommand{\BRpip}{\text{Br}_{\pi}^{\dl}}
\newcommand{\BRDp}{\text{Br}_{D^+}^f}
\newcommand{\Npi}{N_\pi^f}
\newcommand{\ndlndlb}{\prn{n_{\dl}-n_{\dlbar}}}
\newcommand{\BR}[1]{\text{Br} \prn{#1} }
\newcommand{\oth}{\text{other}}

\begin{document}

\title{Making the Universe at 20 MeV}

\author{Gilly Elor}
\email{gelor@uw.edu}
\affiliation{Department of Physics, University of Washington, Seattle, WA 98195, U.S.A.}

\author{Robert McGehee}
\email{rmcgehee@umich.edu}
\affiliation{Leinweber Center for Theoretical Physics, Department of Physics,
University of Michigan, Ann Arbor, MI 48109, USA}
\affiliation{Berkeley Center for Theoretical Physics, University of California, Berkeley, CA 94720, USA}
\affiliation{Theory Group, Lawrence Berkeley National Laboratory, Berkeley, CA 94720, USA}

\begin{abstract} 
We present a testable mechanism of low-scale baryogenesis and dark matter production in which neither baryon nor lepton number are violated. Charged $D$ mesons are produced out-of-equilibrium at tens of MeV temperatures. The $D$ mesons quickly undergo CP-violating decays to charged pions, which then decay into dark-sector leptons without violating lepton number. To transfer this lepton asymmetry to the baryon asymmetry, the dark leptons scatter on additional dark-sector states charged under lepton and baryon number. Amusingly, this transfer proceeds without electroweak sphalerons, which are no longer active at such low scales. We present two example models which can achieve this transfer while remaining consistent with current limits. The required amount of CP violation in charged $D$ meson decays, while currently allowed, will be probed by colliders. Additionally, the relevant decays of charged pions to dark-sector leptons have been constrained by the PIENU and PSI experiments and will be further explored in upcoming experiments. 
\end{abstract}
\maketitle

\section{Introduction}
\label{sec:Intro}
The standard model of inflationary cosmology predicts a Universe born with equal parts matter and anti-matter, necessitating a dynamical mechanism to generate an asymmetry which seeds the complex structures observed today. The required primordial baryon asymmetry of the Universe (BAU) is inferred to be 
\al{
\YBmeas \equiv (n_{B}-n_{\bar{B}})/s = \left( 8.718 \pm 0.004 \right) \times 10^{-11}\,,
} 
from measurements of the Cosmic Microwave Background (CMB)~\cite{Ade:2015xua,Aghanim:2018eyx} and light element abundances after Big Bang Nucleosynthesis (BBN)~\cite{Cyburt:2015mya,pdg}. Discovering \textit{baryogenesis}, the mechanism responsible for generating this asymmetry, is therefore critical to understanding our very existence.

A mechanism of baryogenesis must satisfy the three Sakharov conditions \cite{sakharov};  C and CP Violation (CPV), baryon number violation, and departure from thermal equilibrium. Many mechanisms of baryogenesis have been proposed, including the perennial favorites: \emph{electroweak baryogenesis}~\cite{Kuzmin:1985mm,Cohen:1990py,Cohen:1990it,Turok:1990in,Turok:1990zg,McLerran:1990zh,Dine:1990fj,Cohen:1991iu,Nelson:1991ab,Cohen:1992yh,Farrar:1993hn} and \emph{leptogenesis}~\cite{Fukugita:1986hr}. But, concrete realizations of these mechanisms encounter significant challenges. Electroweak baryogenesis models often predict electric dipole moments of electrons, neutrons, and atoms which are ruled out by experiments~\cite{Andreev:2018ayy}. On the other hand, leptogenesis models typically occur at high scales and involve very massive particles, thereby making experimental confirmation unlikely.\footnote{\footnotesize{See \cite{Dror:2019syi} for an interesting proposal, as well as \cite{Hernandez:2016kel} and references therein for testable seesaw scenarios which achieve baryogenesis.}} Therefore, exploring novel baryogenesis mechanisms is well motivated, especially if they address other outstanding mysteries of the Standard Model of particle physics (SM) and are discoverable in the near-future. 

While the mechanism of baryogenesis is necessary to explain the origin of the complex visible structures we observe today, such structures \emph{only} constitute roughly 5\% of the energy budget of the Universe. The SM does not explain the nature and origin of dark matter (DM), the gravitationally inferred component of matter which makes up roughly 26\% of the energy of the Universe ~\cite{Ade:2015xua,Aghanim:2018eyx}. Experimental searches for DM at colliders and direct detection experiments, together with studies of the possible indirect effects of DM in astrophysical observations, have yet to shed light on its nature. 

\begin{figure*}[t!]
\centering
\hspace{-0.15 in}
\includegraphics[width=0.98 \textwidth]{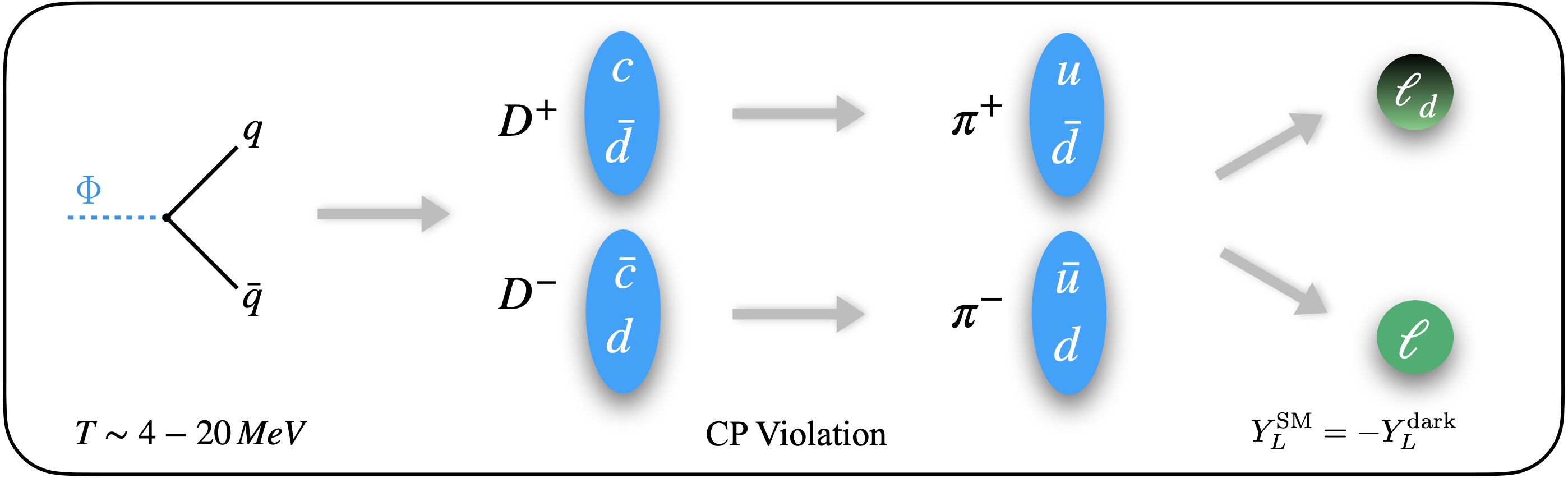}
\caption{Summary of the mechanism by which a lepton asymmetry is produced from late-time production of charged $D^\pm$ mesons. Here we consider CP-violating decays of the $D^\pm$ mesons into final states involving an odd number of charged pions. The charged pions decay into dark- and visible-sector leptons without violating lepton number, producing  equal and opposite visible- and dark-sector asymmetries. }
\label{fig:setup}
\end{figure*}

Many particle physics models have been proposed to explain the nature and origin of DM. However, with the simplest scenarios becoming ever more constrained, richer dark or hidden sectors containing multiple particles with new interactions and symmetries become more interesting.\footnote{\footnotesize{While a rich dark sector may seem less compelling due to its complexity, it is a well motivated scenario from a top-down perspective. Furthermore, the SM displays significant richness, present authors included, despite its meager contribution to the energy budget of the Universe. It would not be too surprising to discover a dark sector with similar complexity.}} Such dark sectors open up a host of new reconstructable cosmological histories \cite{Cheung:2010gj,Chu:2011be,Hochberg:2014dra}, which may be tested by colliders \cite{Elor:2018xku,Tsai:2015ugz,Cheung:2010gk}, direct detection and neutrino experiments \cite{Dror:2019onn,Dror:2019dib, Dror:2020czw,Essig:2015cda,Krnjaic:2017tio,Geller:2020tyv}, and indirect searches \cite{Hooper:2019xss,Elor:2015bho,Barnes:2020vsc}. Moreover, an interesting subset of those models also explain the BAU. For instance, in many models of Asymmetric Dark Matter \cite{Davoudiasl:2012uw,Petraki:2013wwa,Zurek:2013wia,Davoudiasl:2010am}, DM carries a conserved charge whose asymmetry is tied to the BAU in a unified framework that explains both asymmetries (\emph{e.g.},~\cite{Hall:2019rld}, and references therein).

In this work, we explore a novel scenario where a dark-sector state is charged under \emph{lepton} number. Assuming late-time production at temperatures of order 20 MeV, mesons which undergo CP-violating decays may then subsequently have their decay products quickly undergo lepton-number-conserving decays into dark leptons. In this way, an equal and opposite \emph{lepton} asymmetry is generated between the visible and dark sectors. In particular, CP violation in charged $D^{\pm}$ meson decays followed by prompt decays of charged pions to light, MeV-GeV scale (dark) leptons may be used to generate such an asymmetry. Intriguingly, this asymmetry is directly linked to SM observables, making this mechanism testable at current and upcoming experiments (see Fig.~\ref{fig:setup} for a summary).

While a late-time production of a lepton asymmetry may be interesting in its own right, to explain the BAU, the lepton asymmetry must generate a baryon asymmetry. We achieve this by minimally extending the dark sector to include low-scale, dark scattering processes which produce an equal and opposite baryon asymmetry in the dark and visible sectors using the initial lepton asymmetry.\footnote{\footnotesize{For other models which transfer an asymmetry from the dark sector to the SM to realize baryogenesis, see \textit{e.g.}~\cite{Shelton:2010ta,Servant:2013uwa,Hall:2019ank}.}} The SM baryon asymmetry is \emph{Frozen-In} via these dark-sector scatterings. In summary, we present here a novel, testable, mechanism of low-scale baryogenesis and DM production utilizing SM $D^{\pm}$ meson decays at late times, effectively making the Universe as we know it at 20 MeV. In contrast with previous mechanisms such as high-scale leptogenesis, this does not involve lepton- or baryon-number violation and does not require Electroweak sphalerons. 

One of the most remarkable features of this model is the ability to achieve baryogenesis, as well as the production of DM, at such low temperatures. Reasonable assumptions may lead one to conclude that a baryogenesis mechanism, regardless of the source of CP violation, must set the asymmetry by $T \gtrsim 38 \text{ MeV}$ \cite{Kolb:1990vq}. Thus, constructing models of low-scale baryogenesis can be a challenge and there are only a few working examples (see \emph{e.g} \cite{Dimopoulos:1987rk,Cohen:2009tx}). Furthermore, recent proposals for solutions to the gauge hierarchy problem such as \emph{N}naturalness \cite{Arkani-Hamed:2016rle} and cosmological relaxation \cite{Graham:2015cka} require the BAU to be generated at a low scale.

If one holds out hope that the requisite CP violation for baryogenesis exists in the SM, one is also inevitably led to consider mechanisms at such low scales. It is often claimed that there is not enough CP violation within the SM alone to provide for the baryon asymmetry, regardless of the baryogenesis mechanism. However, there are potentially abundant and untapped sources of CP violation in QCD resonances: meson oscillations~\cite{Elor:2018twp,Alonso-Alvarez:2019fym,Alonso-Alvarez:2021qfd} and meson decays, as in this work. Thus, there's a relatively unexplored swath of theory space in which the SM alone provides the necessary CP violation via mesons, allowing for different realizations of \emph{Mesogenesis}. The focus on $D$ mesons in this work is a natural sequel to the $B$ meson work started in~\cite{Elor:2018twp,Alonso-Alvarez:2019fym,Alonso-Alvarez:2021qfd} and predicts a unique set of experimental observables.

This paper is organized as follows. First in Sec.~\ref{sec:mech}, we introduce the mechanism. Next in Sec.~\ref{sec:Details}, we present the details by which baryogenesis is achieved; we solve a set of Boltzmann equations for the lepton and baryon asymmetry and demonstrate that the BAU can be achieved in light of known limits on the CP violation and branching fractions of $D^\pm$ mesons. We also discuss the way in which the correct DM relic abundance can be achieved. Next in Sec.~\ref{sec:DarkSector}, we present two models and demonstrate that they can accommodate a sizeable dark-sector scattering to produce the BAU. We conclude with a discussion of possible extensions, additional variations of \emph{Mesogenesis}, and other future directions in Sec.~\ref{sec:dis}. App.~\ref{app:BoltzDerive} contains a detailed derivation of the Boltzmann Equations. In App.~\ref{app:DecayModes}, we tabulate the relevant $D^\pm$ decay modes and the current limits on their branching fractions and CP asymmetries. 

\newpage
\section{The Mechanism}
\label{sec:mech}
We now introduce the mechanism of baryogenesis and DM from $D^\pm$ mesons. First, we assume the late decay of an inflaton-like scalar field $\Phi$ into quarks and anti-quarks when the temperature of the Universe was roughly tens of MeV. In particular, we assume the decay occurs at temperatures in the range $T_{\rm BBN}  \lesssim  T_R \lesssim T_{\rm QCD}$, so that the produced quarks hadronize but do not spoil the predictions of BBN. $T_R$ is the ``reheat temperature'' corresponding to the time at which $\Phi$ decays. Such a field $\Phi$ may arise naturally out of multi-field inflation models, or may be identified as a flavon in the context of flavor theories. While such models are interesting to consider, for generality, we remain agnostic about the nature of $\Phi$ and simply consider its mass and decay width $(m_\Phi, \Gamma_\Phi)$, as well as relevant branching ratios introduced later.

The produced quarks and anti-quarks hadronize into an equal number of mesons and anti-mesons. By adjusting the mass and decay width of $\Phi$, we consider scenarios in which $D^\pm$ mesons (with mass of $1.87$ GeV) are produced out of equilibrium. Thus, the temperature must be low enough so that $D^\pm$ decay before annihilating with other species. The $D$-meson lifetime is $ \tau_{D} =  1.5\times 10^{9}\,\text{MeV}^{-1}$~\cite{pdg}, while the typical cross section for hadrons is determined by the pion mass $ \sigma  \simeq m_\pi^{-2} \sim \mathcal{O}(10\,\text{mb})$. Following the argument in \cite{Elor:2018twp}, we find an upper bound on the reheat temperature such that the $D^\pm$ mesons decay before annihilating:
\bea
3.5  \,\text{MeV} \, \lesssim  \, T_{\rm R} \, \lesssim \, 20 \,\text{MeV} \,.
\label{eq:TR20MeV}
\eea 
The lower bound of 3.5 MeV comes from the requirement that the asymmetry generation completes before SM neutrino decoupling and we restrict our reheat temperatures to this range \cite{Hasegawa:2019jsa,Kawasaki:1999na,Kawasaki:2000en}.

The $D^\pm$ mesons then undergo CP-violating decays into an odd number of charged pions. Since these decays occur out of equilibrium, an asymmetry in charged pions is temporarily generated. These charged pions themselves quickly decay into a lighter, dark-sector Dirac fermion $\dl$ which carries \emph{visible} sector lepton number ($L=+1$). Since annihilations of pions are subdominant to their decays for the range of temperatures in Eq.~\eqref{eq:TR20MeV}, these fast pion decays are able to happen before any appreciable washout of the temporary pion asymmetry. It may be possible to reheat at slightly higher temperatures than those in Eq.~\eqref{eq:TR20MeV} at the cost of washing out some pion asymmetry as well as depleting some initial $D$ meson abundance. However, calculating a more accurate reheat temperature upper bound would require solving the Boltzmann equations for both charged $D$'s and pions while accounting for non-negligible scattering with the rest of the SM bath. This complexity is beyond the scope of this work and we instead take the conservative upper bound in Eq.~\eqref{eq:TR20MeV}. By introducing this new, dark-sector decay channel for pions, an asymmetry can start to form between the dark and visible sectors. Without it, the generated charged pion asymmetries would wash out.

We consider decays of charged pions into dark and SM leptons that proceed through an effective operator of the form 
\bea
\label{eq:LviolatingOp}
\mathcal{O} = \frac{1}{\Lambda^2} \Bigl[ \bar{d} \Gamma^\mu u \Bigr] \Bigl[\dlbar  \Gamma_\mu \ell \Bigr] \, + \text{h.c.}\, ,
\eea
where $\ell$ is a SM charged lepton and $\Gamma^\mu$ represents all possible distinct Lorentz tensors. The UV model from which the  operator in Eq.~\eqref{eq:LviolatingOp} arises depends on the Lorentz structure. For instance, a scalar operator could arise from a charged scalar mediator similar to \cite{Ibarra:2015fqa}, 
while a vector operator could arise from a new vector of a left-right symmetric model e.g. \cite{Dror:2019dib}. Depending on the UV model, to be consistent with current constraints, the scale $\Lambda$ could be anywhere from hundreds of GeV to a few TeV.

The result of the fast decays,
\bea
\pi^+ \rightarrow \dl + \ell^+ \,, \quad m_{\dl} < m_{\pi^+} - m_\ell \,,
\eea
along with the conjugate decays, is the generation of a lepton asymmetry in the dark sector
\bea
\Ydl \equiv  \left(\frac{n_{\dl} - n_{\dlbar}}{s} \right) \,,
\label{eq:YLdef}
\eea
which is equal and opposite to a lepton asymmetry created in the visible sector. Throughout this work, we use the common co-moving yield variables $Y$ defined as the ratio of the number density to the entropy density in the SM bath. In the absence of any other lepton-charged, dark-sector states, $\Ydl = \YLdark$, the total lepton asymmetry in the dark sector. But, in later sections, we introduce additional dark-sector leptons in order to generate the baryon asymmetry, resulting in $\Ydl \le \YLdark$.\footnote{\footnotesize{In much of the parameter space that results in the measured baryon asymmetry, the dark lepton asymmetry is much greater and $\Ydl \approx \YLdark$ even after baryogenesis completes.}} Regardless, since we never introduce lepton-violating interactions, the following is always true: 
\al{
\YLdark = - \YLSM.
}
In this way, lepton asymmetries are generated in both the dark and visible sectors while \emph{conserving the total lepton number} of the Universe.\footnote{\footnotesize{This mechanism does not require lepton number violation. But the presence of lepton violation, for instance in neutrino masses, will not spoil this mechanism.}}

The generated lepton asymmetry is directly related to SM observables,
\bea
\YLdark   \,\,\, \propto \,\,\, \BRpip \sum_f \ACP \BRDp \ \,,
\label{eq:etaLpion}
\eea
where $\BRpip \equiv \BR{\pi^+ \to \dl + \ell^+}$, the sum is over final states $f$ which contain an odd number of $\pi^\pm$, and $\ACP$ is the CP violation observable for a given decay mode, defined by
\begin{align}
\ACP \,=\, \frac{\Gamma(D^+ \rightarrow f) - \Gamma(D^- \rightarrow \bar{f})}{\Gamma(D^+ \rightarrow f) + \Gamma(D^- \rightarrow \bar{f})} \,.
\label{eq:ACP}
\end{align}
$\BRDp \equiv \BR{D^+ \to f}$ is the branching fraction of the $D^+$ decay (the relevant decay modes and the current limits on their branching fractions and CPV are summarized in Table.~\ref{table:DtoPimodes}). The current limits on $\BRpip$ may be extracted from the limits on sterile neutrinos. Recasting and imposing current limits  for charged pion decays into electrons~\cite{Aguilar-Arevalo:2017vlf,Bryman:2019bjg}, we find that the allowed branching ratio is not large enough to generate the requisite asymmetry when $m_{\dl} > 1 \, \text{MeV}$. However, the branching ratio is unconstrained for sub-MeV $\dl$ masses so that this decay mode can generate the entire asymmetry. Recasting the most current bound from PIENU \cite{Abela:1981nf,Bryman:2019bjg} for final-state muons yields 
\begin{align}
\label{eq:pionlimitsMPIENU}
\nonumber
\text{Br}(\pi^\pm \to \mu^\pm + \text{MET})  \,&\lesssim \,10^{-6}- 10^{-5} \,, \\ 
\quad \text{for} \,\, \, 15.7\, \text{MeV}\, &< m_{\dl}\, < 33.8 \,\text{MeV} \, ,
\end{align}
which is just at the threshold of producing enough asymmetry. 
For lighter $\dl$ masses, constraints can be recast from PSI \cite{Aguilar-Arevalo:2019owf,Bryman:2019bjg} 
\begin{align}
\label{eq:pionlimitsM}
\nonumber
\text{Br}(\pi^\pm \to \mu^\pm + \text{MET})  \,&\lesssim \, 10^{-3}  \,, \\ 
\quad \text{for} \,\, \, 5\, \text{MeV}\, &< m_{\dl}\, < 15 \,\text{MeV} \, .
\end{align}
Note that for $\sim$1-5 MeV, the bound on the branching fraction can be as weak as $10^{-2}$. Given the $\dl$ mass dependence, these bounds do not constrain the entire parameter space of interest to us; as with decays to final-state electrons, sub-MeV $\dl$ masses lead to completely unconstrained branching ratios.  

Improved measurements of these decays will be the focus of upcoming searches at future experiments and as such will be able to further probe this mechanism \cite{privcommDBryman}. In what follows, we will demonstrate that a large lepton asymmetry may be generated which is consistent with current experimental bounds and may be probed in the future.

Baryogenesis is achieved by transferring\footnote{\footnotesize{For simplicity, we refer to this as an asymmetry transfer, since the asymmetry in $\dl - \dlbar$ is being partially translated into an asymmetry in SM baryons and dark-sector particles. Note that the total lepton asymmetry in the dark (and SM) sectors does not change as a result, and so this is not a ``transfer," strictly speaking.}} the dark lepton asymmetry into a SM baryon asymmetry using additional dark-sector states and dynamics which can be rich and possibly reconstructable. In particular, we consider $\dl$ interactions with additional dark-sector states ($\di$ and $\df$) that carry lepton- and baryon-number which can transfer the dark lepton asymmetry into a SM baryon asymmetry. Critically, this dark scattering can occur through an operator which conserves the total baryon and lepton number of the Universe; a dark-sector lepton asymmetry is partially transferred to equal and opposite dark- and visible-sector baryon asymmetries. Schematically, we consider scatterings of the form 
\bea
\label{eq:darkScatt}
\dlbar + \di \rightarrow \df +  \BSM \,,
\eea
where $\BSM$ is a SM baryon, and $\di$ and $\df$ are the gauge-singlet, dark-sector states which may be fermions or scalars depending on the exact dark-sector model. For possible baryon and lepton number charge assignments, see Table~ \ref{table:DarkLandBAll}. Note that the mass of a dark-sector state charged under baryon number must be greater than $1.2$ GeV~ \cite{McKeen:2018xwc}, but dark leptons may be considerably lighter. Additional kinematic and stability requirements will be model dependent, and we leave these details for Sec.~\ref{sec:DarkSector}.

Depending on the details of the dark-sector charge assignment and the UV model, either $\di$ or $\df$ (or both) may constitute (part of) DM. A $\mathbb{Z}_2$ discrete symmetry will generically need to be imposed to stabilize the DM and evade washing out the produced asymmetry. In Sec.~\ref{sec:DarkSector}, we describe the cosmological assumptions and possible models of the dark sector that allow for a large-enough cross section to transfer the asymmetry consistent with current bounds as well as produce the measured DM relic abundance. 

\section{The Details}
\label{sec:Details}
Having given a broad-brush overview of the important ingredients of this mechanism in the previous section, we move on to calculate the relevant matter contents in detail. We consider the generation of the (dark-sector) lepton asymmetry, (visible-sector) baryon asymmetry, and DM in turn.

\subsection{Generating a Lepton Asymmetry}
\label{sec:GenLeptonAsym}
In this section, we demonstrate that a dark lepton asymmetry equal to (or much greater than) the measured baryon asymmetry may be generated via the processes outlined in Fig.~\ref{fig:setup}, postponing a discussion of how it may be transferred to a SM baryon asymmetry to Sec.~\ref{sec:baryonasym}.

In order to numerically solve for the generated lepton asymmetry, we consider the coupled Boltzmann equations which track the production and CP-violating decays of $D^\pm$ mesons into $\pi^\pm$, which then subsequently decay into dark leptons and anti-leptons. For simplicity, we compute the generated lepton asymmetry for the range of reheat temperatures in Eq.~\eqref{eq:TR20MeV} so that annihilations of $D^\pm$ and $\pi^\pm$ mesons can be ignored. The reheat temperature is defined by $4 H \left(T_R\right) = \Gamma_\Phi$, so that Eq.~\eqref{eq:TR20MeV} corresponds to an inflaton decay width in the range $\Gamma_\Phi \in \left[  1 \times 10^{-22} \, \text{GeV} ,\, 3 \times 10^{-21} \, \text{GeV} \right]$. Additionally, as the inflaton must be heavy enough to produce $D^\pm$, its mass must be in the range  $m_\Phi \in \left[ 5\, \text{GeV}\,, \, 100\, \text{GeV}\right]$. $\Phi$ late decays to radiation so that the evolution of the $\Phi$ number density and the radiation density are governed by the interplay of the following Boltzmann equations 
\bea
\label{eq:nPhiBoltz}
 \frac{d n_\Phi}{d t} + 3 H n_\Phi &=& -\, \Gamma_\Phi n_\Phi \,, \\ 
\frac{d \rho_{\rm rad}}{dt} + 4 H \rho_{\rm rad} &=& +\, \Gamma_\Phi m_\Phi n_\Phi \,,
\eea
where the Hubble parameter is given by
\bea
H^2  = \frac{8 \pi}{3 M_{\rm Pl}^2}\left( \rho_{\rm rad} + m_\Phi n_\Phi \right)\,.
\eea
We assume that $\Phi$ was in equilibrium at some high temperature with the bath and as such has a number density $\propto T^3$. While it may be possible to achieve this mechanism in an inflationary model where $\Phi$ is identified as the inflaton, this assumption of high-temperature equilibrium simplifies this analysis at the cost of presuming other scalars responsible for inflation.

Since the focus in this section is on the lepton asymmetry, we assume a minimal dark sector with only $\dl$ and $\dlbar$ and do not include any additional dark-sector states or interactions, deferring this discussion to Sec.~\ref{sec:baryonasym}. Since the formation and subsequent decay of the $D^\pm$ meson and the following decay of the $\pi^\pm$ meson occurs quickly (before any scattering effects can significantly change the abundance of these mesons), the generated dark-sector lepton asymmetry can be written simply as (for a detailed derivation, see App.~\ref{app:BoltzDerive}) 
\bea
\label{eq:BEasymL}
\frac{d}{dt} \ndlndlb &+& 3 H \ndlndlb =  \\
&2& \Gamma^D_\Phi n_\Phi \BRpip \sum_f \Npi \ACPfrac \BRDp,\nonumber
\eea
where $\Npi$ is the number of $\pi^+$ minus the number of $\pi^-$ in each channel labeled by $f$. Note that only decay modes with an odd number of charged pions contribute, as expected. Here we define ${\Gamma_\Phi^D \equiv \Gamma_\Phi \text{Br}(\Phi \rightarrow c) \text{Br} (c \rightarrow D)}$ (where we account for the possibility that $\Phi$ can also populate dark-sector states). Also,  ${\ACPfrac \equiv \ACP/(1+\ACP) \approx \ACP}$ for most decay channels since $\ACP$ is a small number. The sum is over the exclusive rates to each of the final states $f$ listed in Table~\ref{table:DtoPimodes}. In this way, an asymmetry in $\dl$ is generated, as defined in Eq.~\eqref{eq:YLdef}, that is equal and opposite to an asymmetry generated in the visible-sector leptons. This asymmetry is interestingly related to observable CP-asymmetries and branching fractions in SM mesons systems. Critically, note again that the total lepton number of the Universe is actually \emph{conserved}, as we have not introduced any lepton-number-violating interactions. 

\begin{figure}[t]
\centering
\includegraphics[width=\linewidth]{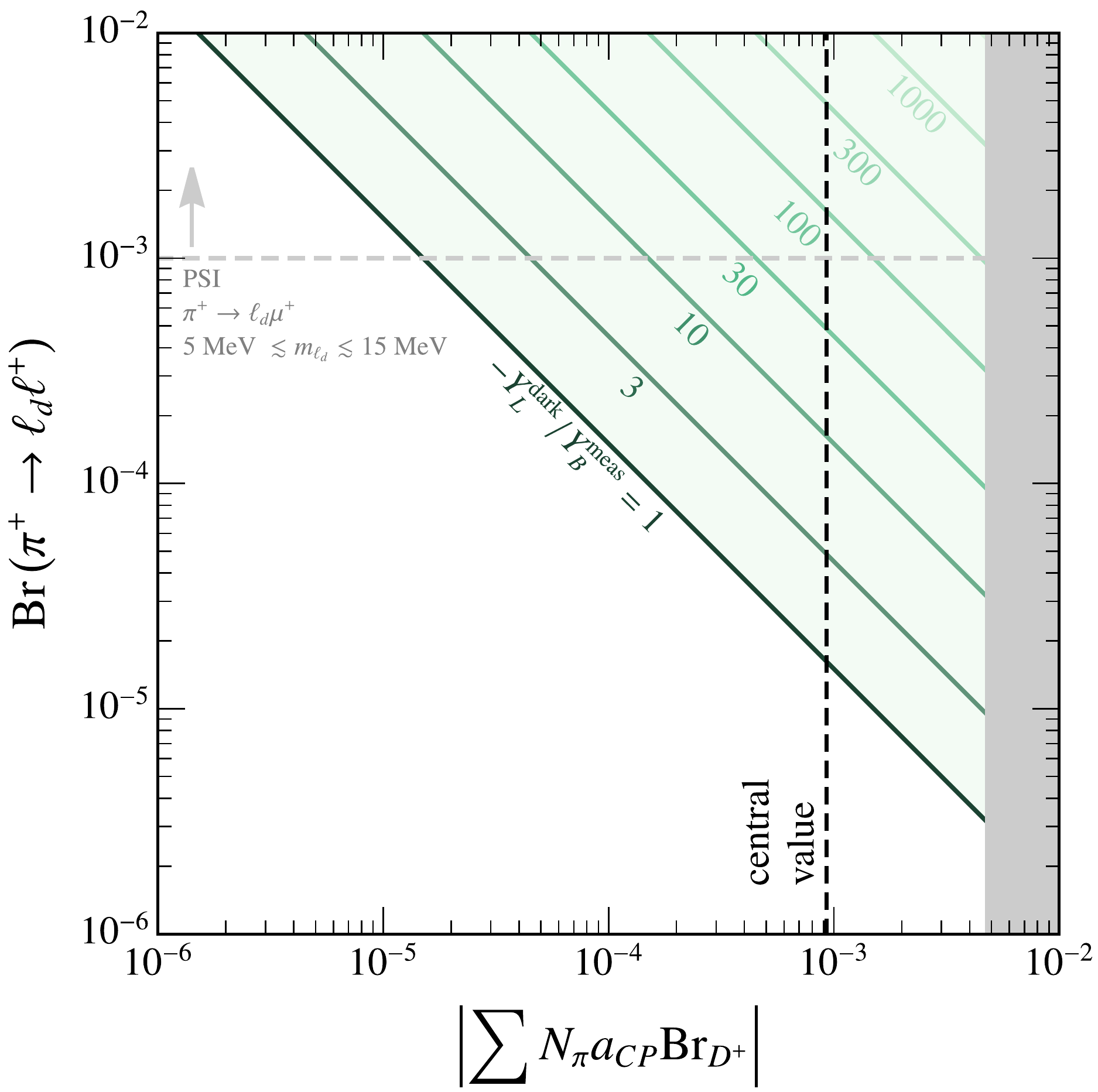} 
\vspace{-0.2 in}
\caption{
The shaded green region corresponds to values of charged $D$ meson and pion observable needed to produce a lepton asymmetry that is equal to or greater than the observed baryon asymmetry $\YBmeas$. The contours correspond to solving the Boltzmann equations Eq.~\eqref{eq:BEasymL}, as summarized in Eq.~\eqref{eq:etaLnum}, with values of $(m_\Phi, T_R) = (5 \, \text{GeV}, \, 20 \, \text{MeV})$ which correspond to maximizing the produced asymmetry. The dotted gray line and the gray shaded region  represents the current limits on the pion branching fraction \cite{Aguilar-Arevalo:2017vlf} and the limits on the sum of the $D$ meson CP asymmetry and branching fraction, respectively.
}
\label{fig:LeptonAsymParamSpace}
\end{figure}
We numerically integrate the above set of Boltzmann equations and float the values of $\sum_f \Npi \ACPfrac \BRDp$, $\BRpip$, $T_{\rm R}$, and $m_\Phi$ to discover the parameter space in which a sizable lepton asymmetry may be generated. We find 
\begin{align}
\frac{\YLdark}{\YBmeas}  \simeq \frac{\BRpip}{10^{-3}} \frac{ \sum_f \Npi \ACPfrac \BRDp}{3\times10^{-5}} \frac{T_{\rm R}}{20\,\text{MeV}} \frac{10\,\text{GeV}}{m_{\Phi}}.
\label{eq:etaLnum}
\end{align}
If all the lepton asymmetry can be instantaneously converted into a baryon asymmetry, then the SM baryon asymmetry will be $\YBSM=\YLdark$. In practice, the dark-sector dynamics need not transfer the asymmetry completely. Therefore, Eq.~\eqref{eq:etaLnum} represents a lower bound on the observables such that baryogenesis can be achieved.  In Fig.~\ref{fig:LeptonAsymParamSpace}, we show contours of $\YLdark / \YBmeas$ for a range of values of the experimental observables $\BRpip$ and $\sum_f \Npi \ACPfrac \BRDp$. Also shown for reference is the PSI constraint from Eq.~\eqref{eq:pionlimitsM} which holds when $5 \,  \text{ MeV}  \lesssim m_{\dl}   \lesssim 15 \text{ MeV}$.

Summing over the relevant $D^\pm$ decay modes in Table~\ref{table:DtoPimodes}, we find 
\bea
\label{eq:Dobsexp}
\sum_f \Npi \ACPfrac \BRDp= \left( -9.3 \times 10^{-4}\right)^{+0.0031 }_{-0.0039}\,,
\eea
where the central value corresponds to taking the central values of both $\ACP$ and $\BRDp$ for each decay channel. The lower bound corresponds to the ``lowest-reasonable'' value for the sum and is calculated in the following way. To make the sum as negative as possible, we take all $\ACP$ values $1\sigma$ below their mean. For channels with values of $\ACP$ which are still positive, we assume their corresponding $\BRDp$ is $1\sigma$ below the mean. For channels which instead (now) have negative $\ACP$, we assume their corresponding $\BRDp$ is $1\sigma$ above the mean. The upper bound in Eq.~\eqref{eq:Dobsexp} is calculated in an analgous way. The measured central value is shown in dashed black in Fig.~\ref{fig:LeptonAsymParamSpace}, while the solid gray region corresponds to the (absolute value) of the most negative possible sum. Comparing Eq.~\eqref{eq:Dobsexp} to Eq.~\eqref{eq:etaLnum}, it is clear that it is possible to generate a dark-sector lepton asymmetry that is orders of magnitude larger than the measured baryon asymmetry.

Future, more precise measurements of $\ACP$ and $\BRDp$ for the various pion decay channels in Table~\ref{table:DtoPimodes} will shift the gray, ruled-out region to the left. Such improvements are expected to be made by experiments such as LHCb. While $\ACP$ are expected to be small in the SM, quantifying them is plagued with the usual technical challenges of the charm sector. If better SM predictions result in $\ACP$ which are smaller than we require for this mechanism, new physics contributions could also enhance $\ACP$ while keeping them within current experimental bounds.

The PIENU experiment has accessed the majority of its data, and as such, an improvement in the sensitivity of $\BRpip$ is unlikely. However, the relevant mass range could be extended as uncertainties are improved which previously made certain areas of phase space difficult to probe. Additionally, next-generation experiments which would improve the limit on the branching fraction are being proposed \cite{privcommDBryman}. Note that for a given UV model generating Eq.~\eqref{eq:LviolatingOp}, the branching ratio $\BRpip$ can be computed and will depend on the scale of the higher-dimensional operator $\Lambda$. This in turn will be constrained by collider and astrophysical searches in a model-dependent way. As a sanity check, we have performed this computation for the charged-scalar mediator scenario e.g. \cite{Ibarra:2015fqa}, and find that direct constraints on the scale $\Lambda$ in this model do not exclude any of the parameter space of Fig.~\ref{fig:LeptonAsymParamSpace}.

\subsection{Generating a Baryon Asymmetry}
\label{sec:baryonasym}
We now complete baryogenesis by elucidating the details by which equal and opposite baryon asymmetries in the dark and visible sectors are frozen-in. We remain agnostic about the dark-sector model which generates the scattering process in Eq.~\eqref{eq:darkScatt}, deferring a detailed discussion to Sec.~\ref{sec:DarkSector}. Instead, we compute how large the cross section must be for the process in Eq.~\eqref{eq:darkScatt} to efficiently transfer the dark lepton asymmetry to the measured baryon asymmetry of the Universe. 

For simplicity, we take $\Phi$ to also decay to $\di$ and therefore require $m_{\di} < m_\Phi$, though another scalar could instead be responsible for this late-time, out-of-equilibrium $\di$ production. The number density of $\di$ therefore evolves according to 
\bea
\label{eq:diBE}
\frac{dn_{\di}}{dt}+3Hn_{\di} &=& \Gamma_\Phi n_\Phi \BR{\Phi \to \di \dibar} \\
&-& \sigv n_{\dlbar} n_{\di}\,, \nonumber
\eea
where $\sigv$ is the thermally averaged cross section\footnote{\footnotesize{While it is technically correct that this is a thermally averaged cross section, the phase space distribution functions will not be the usual thermal Maxwell-Boltzmann distributions. Rather, they are determined by the kinematics of the relevant decays and Hubble expansion.}} of the baryon-transfer process in Eq.~\eqref{eq:darkScatt}. Detailed derivations of all of the Boltzmann equations in this section may be found in App.~\ref{app:BoltzDerive}.

\begin{figure}[t]
\centering
\includegraphics[width=\linewidth]{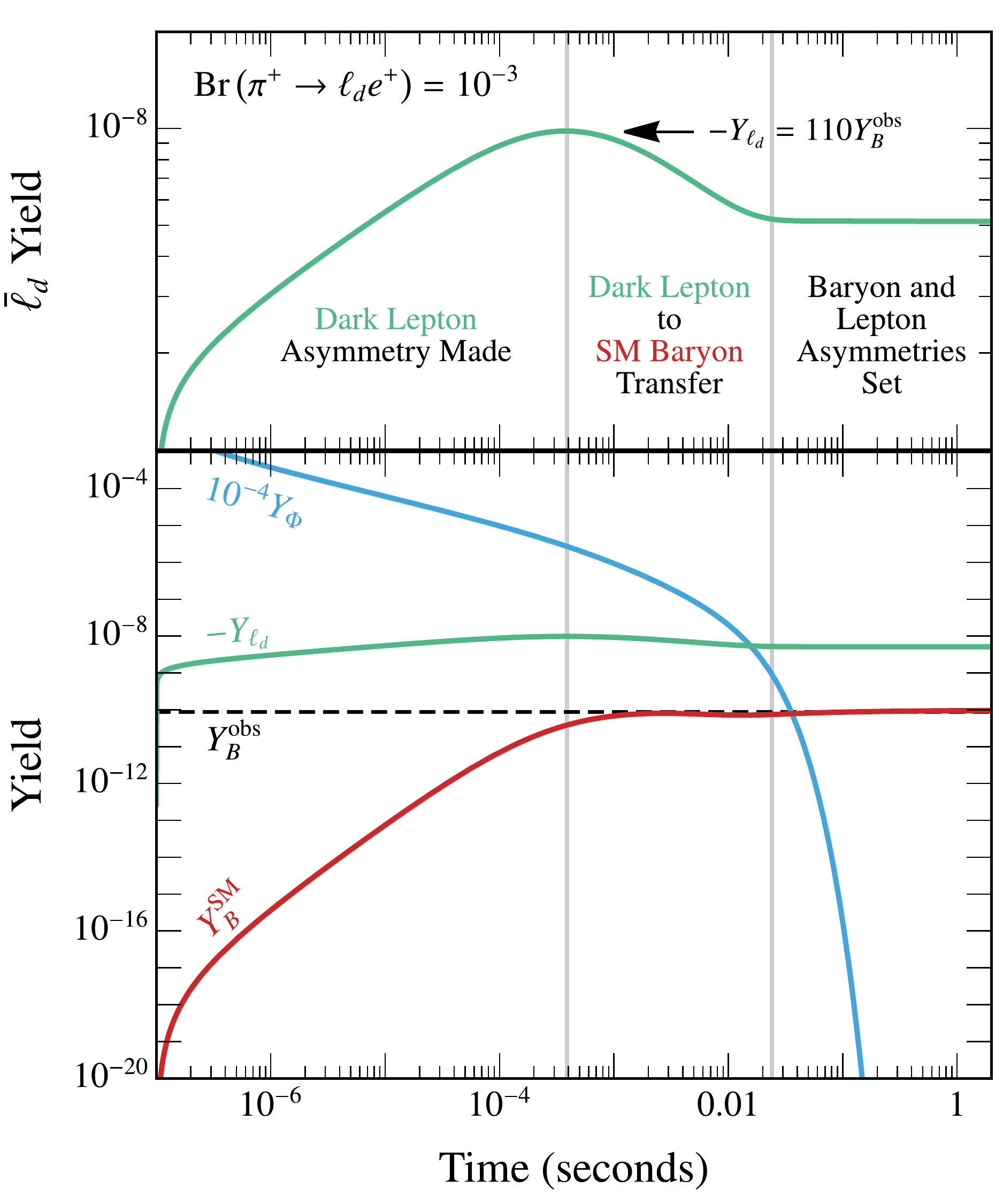} 
\caption{Numerical solutions for the yields of relevant species for a benchmark point which produces the observed baryon asymmetry. We take $T_R=10 \text{ MeV}$, $m_\Phi = 6 \text{ GeV}$, $\BR{\Phi \to \chi_1 \bar{\chi}_1}=0.1$, $\sigv = 1 \times 10^{-15} \text{ GeV}^{-2}$, and $\BRpip=10^{-3}$, and for $\sum_f \Npi \ACPfrac \BRDp$ we take the maximum value in Eq.~\eqref{eq:Dobsexp} (although note that saturating this bound was not required to get the right asymmetry). There are three distinct phases which are delineated by vertical dashed, gray lines, and are highlighted particularly in the top panel. See the text for details.}
\label{fig:BenchmarkBoltz}
\end{figure}

Recall that the evolution of the asymmetry in $\dl$ in Eq.~\eqref{eq:BEasymL} simply tracked the production of a lepton asymmetry. We modify this equation to include the relevant scattering term and obtain the evolution equation for the asymmetry in $\dl$: 
\bea
\label{eq:ldasymwithscatt}
&& \frac{d}{dt} \ndlndlb + 3 H \ndlndlb = \\
&&\qquad 2 \Gamma^D_\Phi n_\Phi \BRpip \sum_f \Npi \ACPfrac \BRDp - \sigv n_{\di} \ndlndlb. \nonumber
\eea
For simplicity,  we take  $n_{\chi_1} \sim n_{\bar{\chi_1}}$ here as both are initially produced in equal amounts from $\Phi$ decays. The Boltzmann equation for the evolution of the SM baryon asymmetry is then given by
\al{
\label{eq:BEbaryonasym}
\frac{d}{dt} \nBSMnBSMb &+ 3 H \nBSMnBSMb = \\
&- \sigv n_{\di} \ndlndlb. \nonumber
}
Next, we turn our attention to numerically solving the set of coupled Boltzmann equations for the baryon asymmetry of the Universe. 

From Eq.~\eqref{eq:BEbaryonasym}, it is clear that the produced lepton asymmetry will be efficiently transferred when the scattering rate $n_{\di} \langle \sigma v \rangle$ dominates over the Hubble expansion. Since we want the transfer to happen quickly, we require the cross section  satisfy
\bea
\frac{n_{\di} \sigv }{H(T)} \Big|_{T = T_R}  \,\, \gtrsim \,\, \frac{\YBmeas}{\YLdark} \,.
\eea
The number density of $\di$ is found by integrating Eq.~\eqref{eq:diBE} and is roughly given by $n_{\di} \sim \BR{\Phi \to \di \dibar} T^3$. We numerically integrate the Boltzmann equations Eqs.~\eqref{eq:nPhiBoltz},~\eqref{eq:diBE}, \eqref{eq:ldasymwithscatt},~\eqref{eq:BEbaryonasym} to solve for lepton and baryon asymmetries, floating the model parameters. We find that the dark scattering cross section is required to be greater than 
\bea
\nonumber
&& \hspace{-0.3in} \langle \sigma v \rangle \,\, \gtrsim \,\,  10^{-16} \, \text{GeV}^{-2} \,  \frac{\YBmeas}{\YLdark} \, \\
&&\,\,\,\, \times \,\,\frac{10 \, \text{GeV}}{m_\Phi}   \frac{20 \, \text{MeV}}{T_R}  \frac{10^{-1}}{\text{Br} (\Phi \rightarrow \di \bar{\di}) }   \,. 
\label{eq:sigmabound}
\eea

In Fig.~\ref{fig:BenchmarkBoltz}, we plot the solution of the Boltzmann equations for a benchmark point that achieves baryogenesis. We plot the yields corresponding to $\Phi$ abundance, $\dl$ asymmetry, and SM baryon asymmetry. There are three clearly distinct regions in the plot, particularly emphasized by the top panel which zooms in on $Y_{\dl}$. First, a dark lepton asymmetry is produced as the inflaton begins to decay. Soon after,  $\dl$-$\di$ scatterings begin to dominate and the $\dl$ asymmetry converts to a baryon asymmetry. As inflaton decay completes, the baryon asymmetry freezes in and the $\dl$ asymmetry is fixed.  

In the following section, we will present possible UV models that can accommodate a  cross section of the size in Eq.~\eqref{eq:sigmabound} while remaining consistent with present constraints. Note that the branching fraction of the inflaton into the dark sector $\BR{\Phi \to \di \dibar}$ depends on the specific inflationary model, but can be sizable. Furthermore, $\sigv n_{\mathcal{B}}$ and $\sigv n_{\df}$ are easily both less than Hubble as these particles are not sourced by $\Phi$. Therefore, any possible washout effects are negligible.

\subsection{Generating the Dark Matter}
Since baryon number is never violated, the measured SM baryon asymmetry is always balanced by an equal and opposite baryon asymmetry in the dark sector. This dark baryon asymmetry, therefore, is always an asymmetric component of DM, and a substantial fraction at that due to the lower bound on baryon-charged masses of 1.2 GeV. Further, there are equal and opposite lepton asymmetries in the dark and visible sectors. If the state(s) which comprise the dark-sector baryons due not also account for this dark-sector lepton asymmetry, then these additional dark leptons must make up a different, asymmetric subcomponent of DM. Clearly, the details depend on baryon- and lepton- number assignments to states $\di$ and $\df$, which we defer to the next section. Here, we just make simple qualitative remarks about how generating the correct DM abundance is relatively straightforward (as compared to generating the baryon asymmetry).

Perhaps the simplest scenario is to assume that the dark baryon-charged state comprises almost the entirety of DM, making it a well motivated case of completely asymmetric DM. In this case, the lightest dark lepton is appreciably lighter than the dark baryon so that it makes up a negligible subcomponent of DM. Thus, $m_{\text{DM}} \sim 5 \text{ GeV}$.

As an alternative, we may also consider the case where the dark baryon state is lighter so that the other dark-sector particles must comprise the remaining relic abundance of DM. These could be new additional states, or just the dark-sector states already present to provide for the baryon-asymmetry transfer. The details here become less relevant to the baryogenesis mechanism considered since there are generic, dark-sector freeze-out possibilities with viable parameter space. 

Both of the above scenarios make one important assumption which we require to be true generically. There must exist a portal between the dark and visible sectors that becomes efficient at late times (before SM neutrino decoupling) to allow the symmetric component(s) of dark-sector states to sufficiently annihilate away, preventing any overabundance of DM or non-negligible contribution to the relativistic degrees of freedom at BBN. The kinetic mixing portal involving a massive dark photon fits our needs here and is a commonly used portal to transfer entropy out of dark sectors at late times (see \emph{e.g.} \cite{Harigaya:2019shz,Hall:2019rld,Tsai:2020vpi} for various such usages).

One additional concern is that protons and anti-protons may not be able to efficiently annihilate below 20 MeV\footnote{\footnotesize{We thank Seyda Ipek for pointing this out.}}, so that a large, symmetric baryon component may freeze-in. One solution is to introduce other dark-sector processes which can efficiently deplete the symmetric component of $\dl$ and $\dlbar$. As a proof of principle, one can imagine the extreme limit where such processes deplete all $\dl$ and leave only the tiny necessary asymmetric amount of $\dlbar$. These would then only freeze-in SM baryons and not anti-baryons, avoiding the problem altogether.

\section{The Models}
\label{sec:DarkSector}

\begin{table}[t!]
\renewcommand{\arraystretch}{2.0}
\setlength{\arrayrulewidth}{.3mm}
\centering
\small
\setlength{\tabcolsep}{13.4pt}
\setlength{\arrayrulewidth}{.3mm}
\begin{tabular}{ |c | c | c || c | c | c | }
    \hline
    Field &  L &   B                 & Field   &  L &  B \\ \hline \hline
     $\di$ &       1   &        0         &  $\di$   &      1    &  1              \\ 
     $\df$ &       0   &       -1      &     $\df$ &       0   &        0            \\ \hline \hline
     $\di$ &       0   &        1        &    $\di$ &       0   &        0          \\ 
     $\df$ &       1   &        0                 &  $\df$ &      -1   &       -1         \\ \hline
\end{tabular}
\caption{Possible baryon and lepton charge assignments for dark-sector states $\di$ and $\df$. Any baron-charged state must be heavier than 1.2 GeV. ({\bf Left}) The two states dark $\df$ and $\di$ involved in the scattering Eq.\eqref{eq:darkScatt} may be charged under SM baryon or lepton number. In this case, DM is multi-component with contributions from $\df$ and $\di$. ({\bf Right}) One of the states involved in the scattering Eq.\eqref{eq:darkScatt} may be a leptobaryon while the other uncharged under SM lepton and baryon number. In this case, the leptobaryon can be the single component of DM.}
\label{table:DarkLandBAll}
\end{table}

Thus far, we have remained agnostic about the nature of the dark-sector fields participating in the baryon-generating process in Eq.~\eqref{eq:darkScatt}. The baryon asymmetry has been computed independently of the details of the dark-sector model and we have found that the dark lepton asymmetry can be efficiently transferred to a SM baryon asymmetry provided the dark-sector cross section is sizable - as given by Eq.~\eqref{eq:sigmabound}. We now turn our attention to the details of the dark sector and possible viable toy models. 

There are several minimal variations of the dark-sector field content which will suffice. The states $\di$ and $\df$ may be identified with a dark baryon and dark lepton as summarized on the left side of  Table~\ref{table:DarkLandBAll}. Alternatively, one state can be neutral and the other a dark leptobaryon\footnote{\footnotesize{We thank Ann Nelson for this suggestion.}}, charged under both SM lepton and baryon number as summarized on the right side of Table~\ref{table:DarkLandBAll}. Note that a dark-sector state carrying baryon number must have a mass greater than $1.2$ GeV in order to be consistent with the observation of old neutron stars \cite{McKeen:2018xwc}, while a dark lepton may be significantly lighter. For concreteness, we enumerate two dark-sector models corresponding to two lepton- and baryon-number charge assignments of $\di$ and $\df$. Depending on the charge assignment, either $\di$, $\df$ or both will constitute (at least part of) DM. 

Regardless of the charge assignment of $\di$ and $\df$, a coupling between a dark-sector baryon and SM fields must be generated. To do so, we simply invoke the model from \cite{Elor:2018twp} and introduce the following interactions which are allowed by all the symmetries:\footnote{\footnotesize{Such models can arise in, for instance, supersymmertic theories \cite{Alonso-Alvarez:2019fym}.}}
\bea
\mathcal{L} \,\,\, \supset  \,\,\,  - y_{u_i d_j} \phi_c^* \bar{u}_i d^c_j - y_{\psi d_k} \phi_c \bar{\psi}_B d_k^c + \text{h.c.}\,.
\eea
Here, $\psi_B$ is a dark-sector Dirac fermion carrying baryon number $B = -1$, and $\phi_c$ is a colored scalar mediator with baryon number $B = -2/3$. Integrating out the heavy $\phi_c$ mediator leads to the following effective \emph{baryon-number conserving} four fermion interaction
\bea
\mathcal{L}_{\rm eff} = \frac{y}{M_{\phi_c}^2} \bar{u}^c_i d_j \bar{d}^c_k \psi_B  \,,
\label{eq:HeffB}
\eea
where we have defined $y \equiv y_{u_i d_j} y_{\psi_B d_k}$.  Note that the colored mediator mass and couplings are constrained to be $M_{\phi_c} \sqrt{y} \gtrsim 1 \, \rm{TeV}$ to be consistent with collider bounds (for details, see \cite{Alonso-Alvarez:2021qfd} and references therein). At low scales, this generates an effective mass mixing between SM baryons and the dark-sector baryon. Note that since $\psi_B$ may couple to protons and neutrons through the operator in Eq.~\eqref{eq:HeffB}, the stability of baryonic matter must be ensured kinematically by $m_{\psi_B} > 1.2  \, \text{GeV}$. The field content is given in Table.~\ref{table:commontoboth}.

Additional dark-sector states are necessary to transfer the asymmetry; the two states $\di$ and $\df$ as well as another mediator. These states are odd under a discrete $\mathbb{Z}_2$ (while $\dl$ and $\psi_B$, which interact directly with the SM, must be even), thereby evading washout and ensuring the stability of the dark-sector lepton and baryon asymmetries. In this way, the dark lepton $\dl$ may scatter 
\bea
\dlbar + \di \rightarrow \df + \bar{\psi}_B \,,
\label{eq:darkScattmodel}
\eea
with charges of $\di$ and $\df $ chosen such that this process conserves baryon, and lepton number. $\bar{\psi}_B$ subsequently mixes into a SM baryon through Eq.~\eqref{eq:HeffB}. Which fields make up the DM depend upon further details of the dark-sector model which we now explore. 

\begin{table}[t]
\renewcommand{\arraystretch}{2.58}
\setlength{\arrayrulewidth}{.3mm}
\centering
\small
\setlength{\tabcolsep}{7pt}
\setlength{\arrayrulewidth}{.3mm}

\setlength{\arrayrulewidth}{.25mm}
\begin{tabular}{ |c || c | c | c  | c  | c |}
    \hline
    Field &        Spin &        L &             B &         $ \,\, \mathbb{Z}_2 \,\,$  &  Mass \\ \hline \hline
   $\phi_c $ &  $0$   &   $0$      &    $-2/3$       & +1    & $\gtrsim \, 1 \, \text{TeV}$  \\ \hline 
   $\dl $ &  $1/2$   &   $1$      &    $0 $       & $+1$     & $\mathcal{O}(10-140 \,  \text{MeV} )$ \\ \hline 
   $\psi_B $ &  $1/2$   &   $0$      &    $ -1$      & $+1$     & $\gtrsim 1.2 \, \text{GeV}$ \\ \hline 
\end{tabular}
\caption{Dark-sector states which interact directly with the SM. $\psi_B$ is a dark-sector baryon introduced in this section to generate interactions between the dark sector and SM baryons through Eq.~\eqref{eq:HeffB}.}
\label{table:commontoboth}
\end{table}

\subsection*{Model 1: DM as Scalar Baryons and Leptons}
In this model, we take $\di$ to have $\prn{L=1,B=0}$ while $\df$ has $\prn{L=0,B=-1}$, corresponding to assignments in the top left two row of Table~\ref{table:DarkLandBAll}. We take both $\chi_{1,2}$ to be scalars and introduce a Dirac fermion mediator $\xi$, which can be MeV scale. We make all three odd under a $\mathbb{Z}_2$. The
DM consists of \emph{both} (asymmetric components of) $\df$ and $\di$, stabilized under the discrete $\mathbb{Z}_2$ symmetry, as to maintain the equal and opposite asymmetries in the dark and visible sectors. 

As discussed above, $\di$ is produced through $\Phi$ decays and $\dl$, generated from the $\pi^\pm$ decay through Eq.~\eqref{eq:LviolatingOp}, then scatters off $\di$ producing $\df$ and the dark-sector fermion $\psi_B$ as in Eq.~\eqref{eq:darkScattmodel}. This scattering is mediated by the dark Dirac fermion $\xi$ and is generated through the following baryon and lepton number conserving Yukawa interactions allowed by the $\mathbb{Z}_2$ symmetry:
\bea
\mathcal{L}  \,\,\, \supset  \,\,\,  y_b  \, \bar{\psi}_B \xi \df + y_l  \,  \dlbar \xi  \di + \text{h.c.} \,.
\label{eq:darkYukawa}
\eea
$\xi$ (which can be relatively light) mediates the s-channel process in Eq.~\eqref{eq:darkScattmodel}. An intermediate $\bar{\psi}_B$ is produced which then quickly mixes into a SM baryon through Eq.~\eqref{eq:HeffB}.  This scattering transfers the dark lepton asymmetry to the SM baryon asymmetry.  

Stability of baryonic matter is ensured kinematically by $m_{\df}> 1.2 \, \text{GeV}$, as with any dark-sector state charged under baryon number. Meanwhile $\di$ can have a sub-GeV mass, unless otherwise restricted by kinematics. In general the Yukawa coupling could induce decays of $\psi_B$ into the dark sector. Since $\psi_B$ transforms into SM baryons via the operator in Eq.~\eqref{eq:HeffB}, we require 
\bea
 m_{\df} + m_\xi > m_{\psi_B}  > m_{\mathcal{B}} \,,
\eea
which also ensures the stability of SM baryons. 

We have computed the thermally averaged cross section corresponding to Eq.~\eqref{eq:darkScatt} for the s-channel scattering in this model and have confirmed that it can easily be sizable enough to satisfy Eq.~\eqref{eq:sigmabound}. We leave a thorough exploration of the corresponding parameter space to future work (and a more detailed UV embedding) and simply present, as a proof of principle, the following result:
\bea
\label{eq:sigmaVModel1}
\langle \sigma v \rangle \,\, \simeq  \,\, && 10^{-15}\, \text{GeV}^{-2} \,  \left( y_l \, y_b \right)^2   \\ \nonumber
&& \times \left(\frac{10 \, \text{MeV}}{m_{\dl}} \right)\left(\frac{20 \, \text{GeV}}{m_{\di}} \right)
\left(\frac{10 \, \text{GeV}}{ m_{\df}} \right) \,.
\eea
Here for simplicity we have taken $m_{\psi_B} \sim 5 \, \text{GeV}$. We have also fixed the dark-sector mediator mass to be $m_\xi = 10$ MeV--- a heavier mediator will result in a slightly smaller cross section. The color mediator mass $M_{\phi_c} \sqrt{y}$ has been set to saturate the collider bound of order 1 TeV. Note that Eq.~\eqref{eq:sigmaVModel1} holds when $m_\Phi \gtrsim m_{\df} + m_{\mathcal{B}}$ and $m_{\Phi} \gtrsim m_{\di}$; as the inflaton populates both $\dl$ and $\di$ in this setup, and the energy available in the scattering will be of order $m_\Phi$.   

The DM will consist of the asymmetric parts of both $\di$ and $\df$. We will  need additional dark-sector interactions to annihilate away any symmetric part of the DM, as it will generically be overproduced, and to obtain the correct relic abundance.  Such a set-up is simple to achieve and there exists a host of dark-sector production mechanisms that can deplete the asymmetry (for instance, see discussion in \cite{Elor:2018twp}). 

\subsection*{Model 2: DM as Fermionic Leptobaryons}
In this second example model, we take $\di$ to have $\prn{L=0,B=0}$ while $\df$ has $\prn{L=-1,B=-1}$, corresponding to charge assignments in the bottom right two rows of Table~\ref{table:DarkLandBAll}. We take both $\chi_{1,2}$ to be fermions and introduce a scalar mediator $\Phi_L$ with $L=1$ which may be light. As in the first toy model, we take all three odd under a $\mathbb{Z}_2$. The $\df$ are leptobaryons in this model and could be, for instance, a neutrino multiplet in a supersymmetric model with an exact $R$-symmetry identified with baryon number, similar to \cite{Alonso-Alvarez:2019fym} (e.g. a right handed sterile neutrio multiplet had a $\nu_R$ with $B=0$, $L = 0$ and $\tilde{\nu}_R$ with $B=-1$, $L=1$). As per Model 1, we have a dark fermionic baryon field $\psi_B$ coupling to the SM by the same UV construction as Eq.~\eqref{eq:HeffB}. 
The following $\Delta B = 0 = \Delta L$ Lagrangian is allowed by all of the symmetries
\bea
\mathcal{L}  \,\, \supset  \,\,  y_b  \, \bar{\psi}_B \df \Phi_L + y_l  \,  \dlbar \di \Phi_L + \text{h.c.} \,.
\eea
This generates the scattering Eq.~\eqref{eq:darkScattmodel} mediated by the dark scalar lepton $\Phi_L$. As with the first model, the scattering cross section in this setup is easily large enough to accommodate baryogenesis. 

In this model, DM is always (partially) comprised of asymmetric $\df$. If $m_{\dl} < m_{\df}$, it will also have an asymmetric component of $\dl$. As in the previous model, we do not illustrate explicitly how to obtain the remainder of the relic abundance, nor do we detail how to remove symmetric components sufficiently. The possibilities here are quite generic to dark-sector DM production and are thus decoupled from the baryogenesis mechanism at hand.

\section{Discussion}
\label{sec:dis}
We have introduced a novel, low-scale mechanism for generating the BAU from the late-time production of mesons. $D^\pm$ mesons decay in a CP-violating way to $\pi^\pm$, which in turn decay into a dark-sector state charged under lepton number. The processes are out-of-equilibrium, occur at tens of MeV, and generate equal and opposite dark- and visible-sector lepton asymmetries. Additional dark-sector states charged under SM baryon number scatter with the dark leptons to transfer this dark lepton asymmetry into the observed SM baryon asymmetry. Since we never explicitly violate lepton or baryon number, these dark states also (partially) comprise DM. 

The matter-antimatter asymmetry is related to experimental observables. The measurement of the CP violation in charged $D^\pm$ decays, $\ACP$, will be improved upon at for instance LHCb. The branching fraction of charged pions into charged leptons and missing energy, which has been constrained by the PIENU experiment, will be further probed at future experiments. 

We have presented two simple, dark-sector models which can efficiently transfer the dark lepton asymmetry and achieve baryogenesis. However, we have remained agnostic about many of the details of the dark sector and have not UV completed these models. Such completions are the subject of future work and will likely open up additional model-dependent, complimentary probes. Another option is to introduce a complex, dark-sector gauge group which transfers the asymmetry through a dark Sphaleron process. We leave the details of this intriguing possibility to future work. 

The study of this mechanism within the context of specific flavor and inflationary models is left to future work. We have focused here on explaining the BAU through $D^\pm$ decays to $\pi^\pm$. Depending on the flavor structure, one could also consider $D^\pm$ decays to Kaons which then decay to dark-sector leptons. Current limits allow for a sizable branching fraction of charged Kaons to muons given by $\text{Br}(K^\pm \to \mu^\pm + X)  <  10^{-3} - 5\times 10^{-6}$~\cite{Artamonov:2014urb,CortinaGil:2017mqf}. Thus, we could repeat the calculations of this work using the currently allowed values for CP-violating decay channels of $D^\pm$ which involve an odd number of Kaons. A benefit of using Kaons is the possibility of reheating at temperatures above 20 MeV since washout-inducing Kaon annihilations stop at higher temperatures than their pion counterparts.

$\Phi$ can also decay into neutral mesons, such as $B^0$, which undergo CP-violating oscillations (this was leveraged in the mechanism of \cite{Elor:2018twp} to explain the BAU). The produced $B^0$ could decay into a dark lepton through Eq.~\eqref{eq:LviolatingOp}, thereby generating a sizable lepton asymmetry of order $Y_L \propto \text{Br} \left( B^0 \rightarrow \text{Mesons} + X \right) \times A_{SL}^{s,d}$, where $A_{SL}^{s,d}$ is the semi-leptonic asymmetry in the $B_{s,d}^0$ systems. Similarly, one could use the CP violation in neutral $D^0$ meson oscillations. Although measurements by LHCb \cite{Aaij:2017urz} have shown there is little evidence of sizable CP violation in $D^0-\bar{D}^0$ mixing, there exist some possible caveats \cite{Nir:2007ac,Golowich:2007ka}. We leave explaining the BAU through new iterations of \emph{Mesogenesis} to future work. 

Generating a lepton asymmetry at low scales can be interesting in its own right, \emph{e.g.} to resolve astrophysical anomalies \cite{Barenboim:2016lxv}. However, applications usually require a larger lepton asymmetry than can be achieved with $D^\pm$ decays alone. To make progress in such a direction, it may be interesting to consider the cumulative effect of multiple CP-violating meson systems (as discussed above). We leave such an avenue to future work. 

It is a well motivated, albeit difficult, exercise for theorists to make the Universe at 20 MeV. That the fingerprints of such an extreme creation may hide within the SM itself compels us to discover whether Nature did it too.

\acknowledgments
We thank Doug Bryman, Tim Cohen, Jeff Dror, Miguel Escudero, David McKeen, Seyda Ipek, Ann Nelson, Jan Stube, Xabier Cid Vidal,  Mark Williams for useful conversations. We thank  Doug Bryman and 
Enrique Fernandez Martinez for pointing out an error in the recast PIENU bound. We thank Miguel Escudero for useful conversations regarding the numerics and recasting sterile neutrino bounds. 
We thank Tim Cohen and Seth Koren for comments on the draft. 
GE is thankful to Ann Nelson for always encouraging the pursuit of creative ideas. GE is supported by the U.S. Department of Energy, under grant number DE-SC0011637. RM was supported in part by NSF grant PHY-1915314 and the U.S. DoE Contract DE-AC02-05CH11231, and is currently supported by the U.S. DoE under grant de-sc0007859. GE thanks the Berkeley Center for Theoretical Physics and Lawrence Berkeley National Laboratory for their hospitality during the completion of this work.

\onecolumngrid
\appendix

\section*{Appendices}

\section{Boltzmann Equations for the Lepton and Baryon Asymmetries}
\label{app:BoltzDerive}
Here we present a detailed derivation of the Boltzmann equations presented in Sec.~\ref{sec:Details}.
It is useful to solve the set of Boltzmann equations in terms of temperature $T$ rather than time. In order to do so, one can write~\cite{Venumadhav:2015pla,Scherrer:1987rr,Hannestad:2004px}
\al{
\frac{dT}{dt} &= \frac{-4 H g_{\star,S}T^4+\frac{30}{\pi^2} \Gamma_\Phi m_\Phi n_\Phi}{T^3 \prn{4g_\star + T dg_\star / dT}}\,,
}
which follows from energy conservation and is valid for energies above the neutrino decoupling temperature $T > 3 \text{ MeV}$, as neutrinos and other light degrees of freedom are still coupled to the plasma. The number of relativistic species $g_\star$ ($g_{\star,S}$) contributing to the energy (entropy) density is given in~\cite{Laine:2006cp}. 

The number density Boltzmann equations for charged $D$ mesons are
\al{
\frac{dn_{D^+}}{dt} + 3 H n_{D^+} &= \Gamma^{D^+}_\Phi n_\Phi - \Gamma_{D^+} n_{D^+}
&\frac{dn_{D^-}}{dt} + 3 H n_{D^-} &= \Gamma^{D}_\Phi n_\Phi - \Gamma_{D^+} n_{D^-},
}
where $\Gamma^{D^+}_\Phi$ is the rate of decay of $\Phi$ to a final state with 1 $D^+$ (we assume no final states with multiple) and $\Gamma_{D^+}$ is the total $D^\pm$ decay rate. We assume that all decays above are much faster than SM annihilations, which is true for $T_R < 20 \,\text{MeV}$. Note that $\Gamma^{D^+}_\Phi=\Gamma^{D^-}_\Phi \equiv \Gamma^{D}_\Phi$ since there is no CP violation in $\Phi$ decays.

The number density Boltzmann equations for the charged pions are:
\al{
\frac{dn_{\pi^+}}{dt} + 3 H n_{\pi^+} &= n_{D^+} \prn{\Gamma_{D^+} \BR{D^+ \to \pi^+ + \oth} +2\Gamma_{D^+} \BR{D^+ \to 2\pi^+ + \oth}} \nonumber \\
&+ \Gamma_{D^-} \BR{D^- \to \pi^+ + \oth} n_{D^-} - \Gamma_{\pi^+} n_{\pi^+}, \\
\frac{dn_{\pi^-}}{dt} + 3 H n_{\pi^-} &= n_{D^-} \prn{\Gamma_{D^-} \BR{D^- \to \pi^- + \oth} +2\Gamma_{D^-} \BR{D^- \to 2\pi^- + \oth}} \nonumber \\
&+ \Gamma_{D^+} \BR{D^+ \to \pi^- + \oth} n_{D^+} - \Gamma_{\pi^+} n_{\pi^-},
}
where $\Gamma_{\pi^+}$ is the total $\pi^\pm$ decay rate. All SM annihilation terms that would appear are negligible relative to the decay terms thanks to the low reheating temperature. The number of different decay terms are due to grouping by the number of final state charged pions, where we are only interested in decays with up to two pions of the same charge (see Tab.~\ref{table:DtoPimodes}). We're also neglecting $\Phi$ decays into lighter quarks which could hadronize into pions. 

For intuition and simplicity, we first solve for the lepton asymmetry generated in the dark sector in the case that the additional dark-sector interactions that give rise to Eq.~\eqref{eq:darkScatt} are absent. Then, the number density Boltzmann equations for the dark leptons are
\al{
\label{eq:elldnoscatter}
\frac{dn_{\dl}}{dt} + 3 H n_{\dl} &= \Gamma_{\pi^+} \BR{\pi^+ \to \ell^+ \dl} n_{\pi^+}, 
&\frac{dn_{\dlbar}}{dt} + 3 H n_{\dlbar} &= \Gamma_{\pi^+} \BR{\pi^- \to \ell^- \dlbar} n_{\pi^-}.
}
We assume any possibly present dark-sector annihilations are slow, $\dl$ is stable, and back scatters of $\ell^+ \dl \to \pi^+$ are slow. 

To find the generated lepton asymmetry in the dark sector, we simply take the difference of the above $\dl / \dlbar $ Boltzmann equations above:
\al{
\frac{d}{dt} \ndlndlb +3 H \ndlndlb = \Gamma_{\pi^+} \BR{\pi^+ \to \ell^+ \dl} \prn{n_{\pi^+}-n_{\pi^-}}.
}
To simplify things analytically, we assume the $\Phi$'s, produced $D$'s, and the subsequently-produced $\pi$'s decay quickly, which is approximately true as long as $H \ll \Gamma_\Phi \ll \Gamma_D \ll \Gamma_\pi$. The right hand side of the above Boltzmann equation becomes
\begin{align}
& \Gamma_{\pi^+} \BR{\pi^+ \to \ell^+ \dl} \prn{n_{\pi^+}-n_{\pi^-}}
 = \Gamma^D_\Phi n_\Phi \BR{\pi^+ \to \ell^+ \dl}  \\
&\qquad \times \,\, \left[ \sum_f \prn{ \BR{D^+ \to 1 \pi^+ + f} - \BR{D^- \to \pi^- + \bar{f}} }  
+2 \sum_f \prn{ \BR{D^+ \to 2 \pi^+ + f} -\BR{D^- \to 2 \pi^- + \bar{f}}  } \right. \nonumber \\ \nonumber
& \hspace{3.6 in} \left. - \sum_f \prn{ \BR{D^+ \to \pi^- + f} - \BR{D^- \to \pi^+ + \bar{f}} } \right] \,.
\end{align}
In the above, the sums over final states $f$ \emph{do not} include any additional charged pions. Rather, the number of charged pions we're considering in a given sum is explicitly highlighted in the decay channel branching ratio. After some partial cancellations between channels with both signs of charged pions, we find that each channel is multiplied by its net `+' charge in $\pi^+$s vs $\pi^-$s, which we define as $\Npi$, and as expected:
\al{
\frac{d}{dt} \ndlndlb +3 H \ndlndlb = \Gamma^D_\Phi n_\Phi \BRpip \sum_f \Npi \prn{ \BR{D^+ \to f}-\BR{D^- \to \bar{f}}}\,.
}

To simplify this further, we rewrite the branching ratio differences in terms of the observable $\ACP$ (as relevant to the $D^\pm$ decays):
\begin{align}
\BR{D^+ \to f}-\BR{D^- \to \bar{f}}&=\BR{D^+ \to f} \frac{2\ACP}{1+\ACP} \equiv 2\BRDp\ACPfrac. \nonumber
\end{align}
Thus, the generated dark-sector lepton asymmetry simplifies to
\al{
\label{eq:psiLasymBE}
\frac{d}{dt} \ndlndlb +3 H \ndlndlb = 2 \Gamma^D_\Phi n_\Phi \BRpip \sum_f \Npi \ACPfrac \BRDp.
}

Next, we modify the story by allowing the interactions in Eq.~\eqref{eq:darkScatt} to cause a net dark baryon asymmetry (and therefore, a net SM baryon asymmetry) to form. With this introduction, notice that $\Ydl \ne \YLdark$ since $\di$ or $\df$ will have lepton number. The $\di$'s and $\dibar$'s are populated by $\Phi$ decays and their Boltzmann equations are
\al{
\frac{dn_{\di}}{dt}+3Hn_{\di}=\Gamma_\Phi n_\Phi \BR{\Phi \to \di \dibar} - \sigv n_{\dlbar} n_{\di}, \nonumber \\
\frac{dn_{\dibar}}{dt}+3Hn_{\dibar}=\Gamma_\Phi n_\Phi \BR{\Phi \to \di \dibar} - \sigv n_{\dl} n_{\dibar},
}
$\df$, $\BSM$, and their conjugates initially have negligible abundances and we assume their abundances are always less than the abundances of $\dl$, $\di$, and their conjugates while these scattering processes are active. We have thus neglected the reverse scattering terms that would contribute to the above. 

The major modification to the previous story (prior to the inclusion of the process in Eq.~\eqref{eq:darkScatt}) is the addition of scattering terms in the Boltzmann equations for $\dl$ and $\dlbar$. Eq.~\eqref{eq:elldnoscatter} becomes 
\al{
\frac{dn_{\dl}}{dt} + 3 H n_{\dl} &= \Gamma_{\pi^+} \BR{\pi^+ \to \ell^+ \dl} n_{\pi^+} - \sigv n_{\dl} n_{\dibar}, \nonumber \\
\frac{dn_{\dlbar}}{dt} + 3 H n_{\dlbar} &= \Gamma_{\pi^+} \BR{\pi^- \to \ell^- \dlbar} n_{\pi^-} - \sigv n_{\dlbar} n_{\di}.
}
Taking the difference, we find that the generated dark-sector lepton asymmetry may now be used to source equal and opposite dark-sector and SM baryon asymmetries:
\al{
\frac{d}{dt} \ndlndlb +3 H \ndlndlb = 2 \Gamma^D_\Phi n_\Phi \BRpip \sum_f \Npi \ACPfrac \BRDp - \sigv n_{\di} \ndlndlb.
}
Note that $n_{\dibar} \approx n_{\di}$ for all times. Finally, the number density Boltzmann equation for the SM baryon asymmetry is simply
\al{
\frac{d}{dt} \nBSMnBSMb + 3 H \nBSMnBSMb = - \sigv n_{\di} \ndlndlb,
}
where again, we have assumed that the backreaction processes are negligible due to the minuscule number densities of $\df$, $\BSM$, and their conjugates relative to $\dl$, $\di$ and their conjugates. 

\newpage
\section{\texorpdfstring{$D$}{D} Meson Decay Modes}
\label{app:DecayModes}
\begin{table*}[t]
\renewcommand{\arraystretch}{2.0}
\setlength{\arrayrulewidth}{.3mm}
\centering
\small
\setlength{\tabcolsep}{0.36 em}
\setlength{\arrayrulewidth}{.25mm}
\begin{tabular}{ |c || c | c |}
    \hline
    $D^{+}$ decay mode & $\ACP/10^{-2}$ & $\BRDp/10^{-2}$ \\ \hline \hline
    $K_S^0 \pi^+$ & $-0.41 \pm 0.09$ & $1.562 \pm 0.031$ \\ \hline  
    $K^- \pi^+ \pi^+$ & $-0.18 \pm 0.16$ & $9.38 \pm 0.16$ \\ \hline  
    $K^- \pi^+ \pi^+ \pi^0$ & $-0.3 \pm 0.6 \pm 0.4$  & $5.98 \pm 0.08 \pm 0.16^\ast$ \cite{Dobbs:2007ab}\\ \hline 
    $K_S^0 \pi^+ \pi^0$ & $-0.1 \pm 0.7 \pm 0.2$ & $6.99 \pm 0.09 \pm 0.25^\ast$ \cite{Dobbs:2007ab}\\ \hline
    $K_S^0 \pi^+ \pi^+ \pi^-$ & $0.0 \pm 1.2 \pm 0.3$ & $3.122 \pm 0.046 \pm 0.096^\ast$ \cite{Dobbs:2007ab}\\ \hline
    $\pi^+ \pi^0$ & $2.4 \pm 1.2$ & $\prn{1.247 \pm 0.033} \times 10^{-1}$ \\ \hline
    $\pi^+ \eta$ & $1.0 \pm 1.5$ & $\prn{3.77 \pm 0.09} \times 10^{-1}$ \\ \hline
    $\pi^+ \eta'(958)$ & $-0.6 \pm 0.7$ & $\prn{4.97 \pm 0.19} \times 10^{-1}$ \\ \hline
    $K^+ K^- \pi^+$ & $0.37 \pm 0.29$ & $\prn{9.35 \pm 0.17 \pm 0.24}^\ast \times 10^{-1}$ \cite{Dobbs:2007ab}\\ \hline
    $\phi \pi^+$ & $0.01 \pm 0.09$ & $\prn{5.70 \pm 0.05 \pm 0.13} \times 10^{-1}$ \\ \hline
    $a_0(1450)^0 \pi^+$ & $-19 \pm 12^{+8}_{-11}$ & $4.5^{7.0}_{-1.8} \times 10^{-2}$ \footnote{\footnotesize{this only includes the subsequent decay mode in which $a_0(1450) \to K^+ K^-$}} \\ \hline
    $\phi(1680) \pi^+$ & $-9 \pm 22 \pm 14$ & $4.9^{+4.0}_{-1.9} \times 10^{-3}$ \footnote{\footnotesize{this only includes the subsequent decay mode in which $\phi(1680) \to K^+ K^-$}} \\ \hline
    $\pi^+ \pi^+ \pi^-$ & $-1.7 \pm 4.2$ & $\prn{3.27 \pm 0.18} \times 10^{-1}$ \\ \hline
\end{tabular}
\vspace{5mm}
\caption{Summary of $D^{+}$ decay modes which violate $CP$ and involve an odd number of $\pi^\pm$ and therefore help to generate a dark-sector lepton asymmetry. 
}
\label{table:DtoPimodes}
\end{table*}
We summarize the relevant $D^+$ decay modes, including their values of $\ACP$ and branching ratios, in Table \ref{table:DtoPimodes}. All quoted values come from the latest Particle Data Group (PDG) \cite{Zyla:2020zbs}, with the following exception. There are some decay modes for which PDG does not provide their own fit to the branching ratio, denoted by an asterisk on the branching ratio value. For these, we use the top listed reference within PDG and cite it in our table accordingly.

\newpage

\twocolumngrid
\bibliography{Refs}
\end{document}